\documentclass[fleqn,twoside]{article}
\usepackage{espcrc2}
\usepackage{graphicx,dcolumn}
\newlength{\colw}
\newcolumntype{d}{D{.}{.}{-1}}

\title{Static-light matrix elements on a dynamical anisotropic lattice}

\author{K. Jimmy Juge\address[TCD]{TrinLat Collaboration, 
        School of Mathematics, Trinity College, Dublin 2, Ireland}, 
        Richie Morrin\addressmark[TCD], Alan \'O Cais\addressmark[TCD], Mike Peardon\addressmark[TCD], 
        Sin\'ead M. Ryan\addressmark[TCD]\thanks{Talk presented by S.Ryan}, Jon-Ivar Skullerud\addressmark[TCD]
}
       
\begin{document}

\begin{abstract}
The static-light matrix element needed to determine $f_B$ is studied on an anisotropic lattice with $N_f=2$. 
The improvement in precision due to stout links and all-to-all propagators is investigated.
\end{abstract}

\maketitle

\section{INTRODUCTION}
The determination of heavy-light decay constants and form factors relevant to CKM phenomenology is a priority for 
lattice QCD. 
However, to make a meaningful contribution the statistical and systematic uncertainties in these calculations 
must be reduced to a few percent for simulations carried out in the full theory. This is extremely costly,
and new methods to improve precision while keeping computational costs modest are important. We report preliminary 
results from our investigations of heavy-light physics on anisotropic lattices using stout links and 
all-to-all propagators. 
\section{SIMULATION DETAILS}
The gauge action used in the simulations described here is Symanzik-improved with a negative adjoint term. It has 
been used previously in a study of glueballs~\cite{MorningstarPeardon} and in an exploratory study of quenched 
heavy-quark physics on anisotropic lattices~\cite{Foley}. The fermion action is designed specifically 
for large anisotropies when the temporal lattice spacing $a_t$ is much finer than the spatical lattice spacing, $a_s$. 
It incorporates the usual Wilson term in the temporal direction and a Hamber--Wu term 
in the spatial directions to remove doublers. The action has been discussed in detail in Refs.~\cite{Foley,SRLat03} 
and is written 
\begin{eqnarray}
  S^\prime &=& \bar{\psi^\prime}M_r\psi^\prime -\frac{ra_t}{2}\bar{\psi^\prime}
               \left( D_t^2 - \frac{g}{2}\epsilon_iE_i\right)\psi^\prime  \nonumber \\
           &&  + sa_s^3\bar{\psi^\prime}\sum_iD^4_i\psi^\prime 
\end{eqnarray}
where the prime indicates that fields have been rotated, $M_r=\mu_r\gamma_iD_i + \gamma_0D_t + \mu_rm_0$, 
$\mu_r=(1+1/2a_tm_0)$ and the spatial derivative is improved 
with a two-hop term. The usual Wilson parameter is $r=1$ and the Wilson-like parameter $s=1/8$. 
To reduce the coupling to UV gluon modes we use stout links~\cite{MP-stout}. 
The preservation of differentiability with respect to the link 
variables of this smearing technique means that the hybrid Monte Carlo algorithm can be straightforwardly applied. 
Quantum fluctuations renormalise the anisotropy $\xi =a_s/a_t$, and the parameters 
$\xi_g$ and $\xi_q$ in the gauge and quark actions respectively must be tuned such that the 
measured anisotropy takes its required value. In the quenched theory the gluon and quark actions can be 
separately tuned. However, in unquenched simulations this is no longer true and the quarks and gluons must 
have a common anisotropy to recover the correct continuum limit of the theory. 
The non-perturbative tuning of this action is under investigation. We require that the anisotropy measured from 
the sideways potential agrees with that determined from the low-momentum relativistic pion dispersion relation. 
This simultaneous tuning requires three initial simulations to find the self-consistent point~\cite{mjp-cairns}. 
The results presented here are based on one (untuned) simulation. The results do not therefore have a physical 
interpretation; the 
point of interest is rather the precision which can be reached using anisotropic lattices and all-to-all propagators. 
\section{RESULTS}
In this exploratory simulation we use the hybrid Monte Carlo algorithm with $N_f=2$ and degenerate sea 
and valence quark masses. We have generated 250 gauge configurations
on a $8^3\times48$ lattice, with $m_\pi/m_\rho\sim0.65$.  The
parameter values are given in Table~\ref{tab:latt-details}.
\begin{table}[t]
\caption{Simulation parameters for one of three runs needed to tune the anisotropy. 
$u_s, u_t$ are the tadpole-improvement coefficients in the spatial and temporal directions.}
\label{tab:latt-details}
\begin{tabular*}{\colw}{@{\extracolsep{\fill}}@{ }ld||ld}
\hline
\multicolumn{2}{l}{Gauge Action} & \multicolumn{2}{l}{Quark Action} \\
\hline
$\beta$ & 1.51  &  $a_tm_0$  & -0.057 \\
$\xi_g$ & 8.0   &   $\xi_q$   & 6.0 \\
$u_s$ & 0.752121&   $u_s, u_t$ & 1.0 \\
\hline
\end{tabular*}
\end{table}

\begin{figure}[th]
\includegraphics[width=\colw]{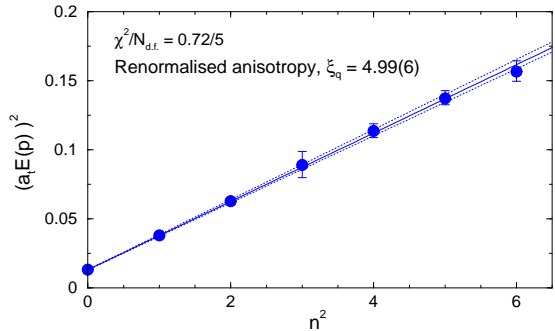}\vspace{-.5cm}
\vspace{-4ex}
\caption[fig:pi-disp]{Dispersion relation for the light quark in the simulation;
  $a_tm_q= -0.057$, close to the strange quark mass. }
\label{fig:pi-disp}
\end{figure}

The combination of anisotropic lattices and stout links allows us to reach momenta, $p>1$GeV. 
Figure~\ref{fig:pi-disp} shows the light pseudoscalar meson dispersion relation obtained from point propagators and 
Jacobi smeared quark fields.
The quark mass is close to the strange quark and the point $n^2=6$ corresponds to momentum $(1,1,2)$ in units of
 $2\pi /a_sL$ or $p\sim 1.5$GeV. 
This result is encouraging since it implies that high-momentum light pseudoscalars can be reliably simulated on 
these dynamical lattices. This is relevant for studies of semileptonic decays such as $B\rightarrow\pi\ell\nu$ 
which have in the past been restricted to the low-momentum regime. 
We have not explicitly disentangled the benefits of the anisotropic lattice from the 
stout links but experience in the quenched case without stout links, indicates that spatial momenta higher than 
$(2,0,0)$ are difficult to resolve, implying that stout links play an important r\^ole in these high-momentum 
simulations. 
\subsection{All-to-all propagators}
A new algorithm to estimate all-to-all propagators was presented at this conference~\cite{jimmyalan04}. With 
all-to-all propagators much more information can be extracted from a gauge configuration than with point sources, 
which is of particular relevance for expensive dynamical simulations. The algorithm discussed in 
Ref.~\cite{jimmyalan04} uses low-mode dominance corrected with a stochastic estimator which yields the exact 
all-to-all propagator in a finite number of quark inversions. Once again we use stout link smearing on the gauge 
fields and Jacobi smearing on the (light) quarks.  
\subsection{Matrix elements}
\begin{figure}[ht]
\includegraphics[width=\colw]{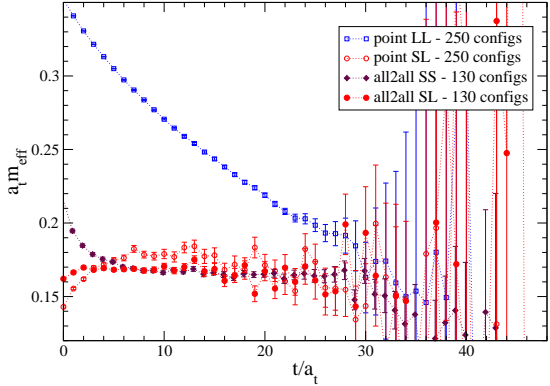}\vspace{-5ex}
\caption[fig:Bmass-all2all]{Effective mass of the static-light meson. The light quark is close to the 
  strange quark mass. Open symbols are determined with point propagators and 
  the solid points with all-to-all propagators.  The all-to-all data
  at large $t/a_t$ have been hidden for the sake of clarity.}
\label{fig:Bmass-all2all}
\end{figure}

We now study the effect of these all-to-all propagators on the static-light correlator. Although relatively cheap 
to calculate, this is known to be noisy and difficult to determine with precision. 
In addition, using all-to-all propagators in the static approximation means 
that $f_B$ can be extracted from a combination of smeared-local (SL) and smeared-smeared (SS) correlators 
rather than the restriction to local-local (LL) and SL correlators when point sources are used. 
The advantages are clearly illustrated in Figure~\ref{fig:Bmass-all2all}. 
The plot shows the effective mass of a heavy-light meson determined using both point and all-to-all propagators. 
The heavy quark was simulated in the static approximation and the light quark is close to 
the strange quark mass. It is important to note that the all-to-all dataset consists of 130 configurations 
whereas the equivalent statistical precision using point sources requires 250 configurations. 

\begin{figure}[ht]
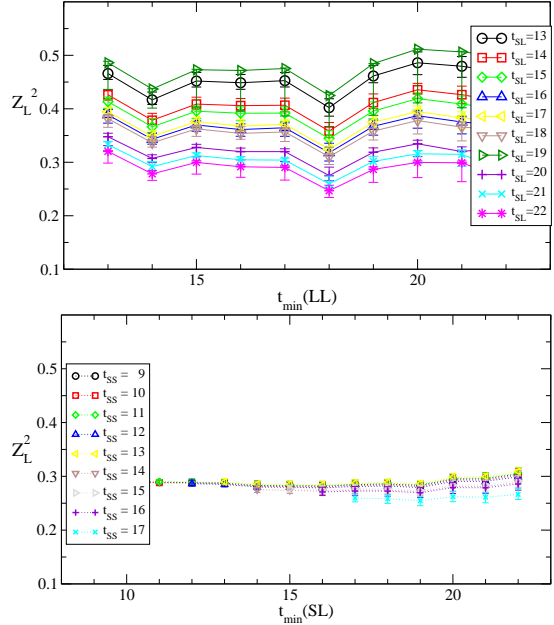
 
\includegraphics[width=\colw]{Zlpt_stab.eps}\\
\includegraphics[width=\colw]{Zl_stab.eps} \vspace{-5ex}
\caption[fig:compare-Zl]{A comparison of the stability of the amplitude with respect to $t_{min}$ for 
  point (upper panel) and all-to-all (lower panel) propagators. }
\label{fig:compare-Zl}
\end{figure}

Using point sources the effective mass and amplitude (which yields $f_B$) is determined 
from a combination of the LL and SL data in a region where they overlap. 
Figure~\ref{fig:Bmass-all2all} shows that this is true only for a few timeslices but also that the purely local 
data in that region are becoming very noisy, decreasing the reliability and precision with which $f_B$ 
can be extracted. 
In contrast, the all-to-all datasets agree over a much larger number of timeslices reducing the 
uncertainties in the fitted mass and amplitude. 
Figure~\ref{fig:compare-Zl} shows a comparison of the matrix-element amplitude determined from point and 
all-to-all propagators. 
These are essentially double sliding window plots. The vertical spread
represents the systematic error in the extracted value of $Z_L^2$ as the
fit range is varied for the point LL and SL correlators and for the
all-to-all SL and SS correlators respectively. The plots are on the same scale and 
the all-to-all data are very impressive especially since these
required only half the number of configurations compared to the 
point sources.
\section{SUMMARY}

The static-light matrix element was computed using all-to-all
propagators for the light quark on an anisotropic (but untuned)
lattice with $N_f=2$, using a stout-link background.  The great
advantage of being able to use smeared-smeared correlators was
demonstrated by a direct comparison with the effective mass taken from
conventional point propagators.  The tuning of the improved, dynamical
anisotropic action is currently being carried out, and the resulting
parameters will be used for a determination of $f_B$.
Work is also in progress to study radially excited S-waves and
P-waves; the preliminary results of this are highly encouraging.


\section{ACKNOWLEDGEMENTS}
KJJ, RM and AOC are supported by the IITAC PRTLI initiative. SR and JIS acknowledge support from IRCSET.

\end{document}